\documentclass[9pt,conference]{IEEEtran}
\usepackage{dcase2026}

\usepackage{bm} 
\usepackage{xspace} 

\usepackage{amsmath,graphicx,url,times,booktabs,tabularx,xcolor}

\newcommand{\RR}{\mathbb{R}}
\newcommand{\IoU}{\textrm{IoU}\xspace} 
\newcommand{\AMR}{Audio Moment Retrieval\xspace}

\title{Overview and Meta-Analysis of DCASE 2026 Challenge Task 6:\\
Audio Moment Retrieval from Long Audio}

\name{Hokuto Munakata$^{1}$,
      Tatsuya Komatsu$^{1}$,
      Keisuke Imoto$^{2}$,
      Taichi Nishimura$^{3}$,
      Huang Xie$^{4}$,
      Tuomas Virtanen$^{5}$
}
\address{
$^{1}$LY Corporation, Japan, \;
$^{2}$Kyoto University, Japan, \;
$^{3}$Sony Interactive Entertainment, Japan, \\
$^{4}$University of Helsinki, Finland, \;
$^{5}$Tampere University, Finland
}

\renewcommand{\baselinestretch}{0.990}

\begin{document}

\maketitle

\begin{abstract}
This paper presents an overview of the Detection and Classification of Acoustic Scenes and Events (DCASE) 2026 Challenge Task 6, Audio Moment Retrieval (AMR) from Long Audio.
Given a several-minute-long audio recording and a free-form text query, AMR aims to retrieve temporal moments in the recording that match the query, where each moment is represented by a pair of start and end timestamps.
This task requires effective cross-modal alignment and long-range temporal modeling.
We describe the task definition, the evaluation metrics, the development and evaluation datasets, and a baseline system that combines a pre-trained MS-CLAP feature extractor with a Detection Transformer (DETR)-based moment-detection network.
On the development data, the baseline trained on a manually annotated dataset and a synthetic dataset achieved Recall1@0.7 of 13.56\%, indicating that AMR in long audio remains a challenging problem.
The challenge attracted 21 teams, which submitted 59 systems in total. The three best systems achieved Recall1@0.7 of 48.59\%, roughly 3.5 times the baseline score.
The results show that strengthening the audio-text feature extractor and the moment-detection network led to substantial performance improvements. Furthermore, the top three teams boosted performance by applying confidence score calibration or ensembling across different temporal resolutions of features.
\end{abstract}


\section{Introduction}
\label{sec:intro}
\begin{figure}[t]
    \centering
    \includegraphics[width=\columnwidth]{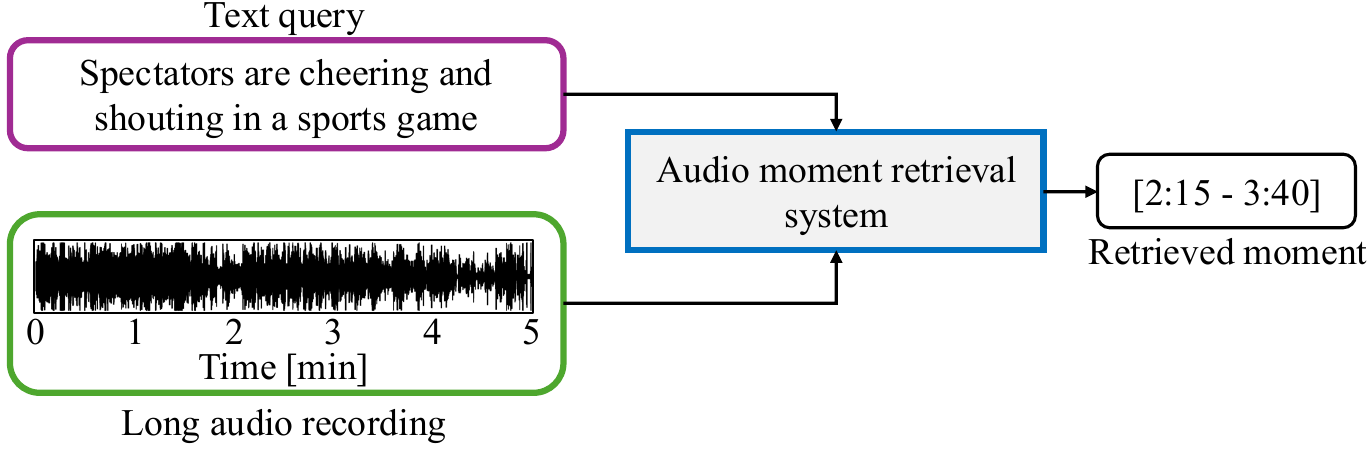}
    \vspace{-6mm}
    \caption{An overview of audio moment retrieval. Given an audio recording and a text query, the system searches for the moment relevant to the query. For the primary metric of this challenge, the model is evaluated by a single predicted moment with the highest confidence score.}
    \label{fig:overview}
\end{figure}

\AMR (AMR) refers to the task of retrieving specific temporal moments within a
long audio recording that align with a given complex free-form text query~\cite{munakata2025amr}.
For example, given a several-minute-long audio recording and a query such as
``\textit{Spectators are cheering and shouting in a sports game}'', a system is
expected to output the start and end timestamps of the moments that match the
query, as shown in Fig.~\ref{fig:overview}.
Here, a \emph{moment} is a contiguous interval defined by its start and end timestamps. A single recording may contain one or several moments that match the query.

Long, untrimmed recordings such as meeting archives, lifelogs, broadcast media, podcasts, and acoustic monitoring are increasingly common, and manually searching for the portion of interest by listening is time-consuming and does not scale.
AMR addresses this by allowing users to describe the target content in free-form text and returning the corresponding moments directly, which helps in indexing and navigating large audio archives.

AMR is positioned among several neighboring tasks.
Clip-level language-based audio retrieval ranks short, pre-segmented clips by
their relevance to a query~\cite{lbar,clap2023,msclap}.
This task assumes the audio is already segmented and does not detect \textit{when} the queried content occurs, whereas AMR handles long and untrimmed recordings, and performs temporal detection.
Sound event detection estimates temporal activities of a predefined, closed set of event classes~\cite{sed}, whereas AMR assumes an open-vocabulary setting specified by a given free-form query, which may describe several overlapping events, or a broader acoustic scene.
The closest equivalent to AMR is video moment retrieval, where Detection Transformer
(DETR)~\cite{detr} models such as Moment-DETR~\cite{momentdetr},
QD-DETR~\cite{qddetr}, UVCOM~\cite{uvcom}, and CG-DETR~\cite{cgdetr} regress the start and end timestamps directly.
AMR transfers this paradigm to the acoustic domain, where temporal cues must be inferred from sound alone.

The initial study proposed the Audio Moment DETR (AM-DETR) baseline together with the large-scale synthetic
Clotho-Moment dataset~\cite{munakata2025amr}, later complemented by the manually annotated real-recording dataset CASTELLA~\cite{castella}.
The central challenge of AMR is to capture long-range temporal context while aligning it with the query using strong audio-text representations, despite the scarcity of annotated long audio data.
Previous work~\cite{castella} has shown that pre-training on synthetic data and fine-tuning on annotated data substantially improve accuracy.

DCASE 2026 Challenge Task 6 is the first DCASE task to address AMR, extending
clip-level language-based audio retrieval to temporal detection.
The organizers provide the datasets, a baseline system, and an evaluation
protocol, so that participants can focus on modeling.
Participants were encouraged to develop advanced audio-text models, to design
networks that effectively capture temporal structure, and to generate or augment
synthetic data to improve training.
In total, 21 teams submitted 59 systems. The three best-performing systems achieved Recall1@0.7 of
$48.59\%$.
While simply replacing the audio-text feature extractor and moment-detection network with more powerful versions led to score improvements, the top-performing teams additionally implemented post-processing techniques such as confidence score calibration and ensembling.

\section{Task definition}
\label{sec:tasksetting}

This section presents the formal definition, notation, and evaluation metrics of
DCASE 2026 Challenge Task 6.

\subsection{Problem formulation and notation}
\label{ssec:formulation}

The input to an AMR system is a long audio recording $\bm{x} \in \RR^{L}$ of
duration $L$ samples and $T$ seconds, and a free-form text query denoted by $q$.
The system outputs a set of $N'$ predicted moments,
\begin{equation}
  \label{eq:output}
  \hat{\mathcal{M}}
  = \bigl\{ (\hat{t}^{\mathrm{s}}_1, \hat{t}^{\mathrm{e}}_1, \hat{c}_1),
           \ldots,
           (\hat{t}^{\mathrm{s}}_{N'}, \hat{t}^{\mathrm{e}}_{N'}, \hat{c}_{N'}) \bigr\},
\end{equation}
where $0\leq \hat{t}^{\mathrm{s}}_n < \hat{t}^{\mathrm{e}}_n\leq T$ denote the start and end
timestamps (in seconds) of the $n$-th predicted moment and $\hat{c}_n \in \RR$ is its confidence score produced by the system.
The predicted moments are ranked by their confidence scores.

The goal of this challenge is to accurately predict the timestamps for moments that match the given query and to assign them high confidence scores.
The query $q$ used in this challenge is relevant to a set of $N$ moments
$\mathcal{M} = \{ (t^{\mathrm{s}}_n, t^{\mathrm{e}}_n) \}_{n=1}^{N}$, where $N \geq 1$, and a single recording may also be associated with multiple queries.
Participants are required to submit at least one moment per query and may submit
multiple moments in descending order of confidence.

\subsection{Evaluation metrics}
\label{sec:metrics}

The correctness of a retrieved moment is determined by its temporal intersection over union (\IoU) with the reference moments.
For a predicted moment $\hat{m} = (\hat{t}^{\mathrm{s}}, \hat{t}^{\mathrm{e}})$
and a reference moment $m = (t^{\mathrm{s}}, t^{\mathrm{e}})$,
\begin{equation}
  \label{eq:iou}
  \IoU(\hat{m}, m)
  = \frac{\lvert \hat{m} \cap m \rvert}{\lvert \hat{m} \cup m \rvert},
\end{equation}
where $\lvert \cdot \rvert$ denotes the temporal length of the (intersected or
unioned) interval.
A predicted moment is regarded as correct when its \IoU with a reference moment is at least a threshold $\theta$.

Systems are evaluated with two ranking metrics, Recall1@$\theta$ and mean
average precision (mAP@$\theta$).
Let $Q$ denote the set of evaluation queries. For a query $q \in Q$, let
$\hat{m}^{(1)}_q$ be its highest-confidence predicted moment and $\mathcal{M}_q$ be the set of reference moments.
Recall1@$\theta$ considers only $\hat{m}^{(1)}_q$ and measures the fraction of
queries whose highest-confidence moment attains $\IoU \geq \theta$ with a reference
moment:
\begin{equation}
  \label{eq:Recall1}
  \text{Recall1@}\theta
  = \frac{1}{|Q|} \sum_{q \in Q}
    \mathbf{1}\!\left[ \max_{m_q \in \mathcal{M}_q}
      \IoU(\hat{m}^{(1)}_q, m_q) \geq \theta \right],
\end{equation}
where $\mathbf{1}[\cdot]$ is the indicator function.
mAP@$\theta$ instead accounts for all predicted moments. For each query, the
predictions are ranked by confidence and matched one-to-one to the reference
moments at threshold $\theta$, yielding an average precision
$\mathrm{AP}_q(\theta)$. mAP@$\theta$ is its mean over queries,
\begin{equation}
  \label{eq:map}
  \text{mAP@}\theta = \frac{1}{|Q|} \sum_{q \in Q} \mathrm{AP}_q(\theta).
\end{equation}
Following~\cite{munakata2025amr}, the averaged mAP further averages mAP@$\theta$
over $\theta \in \{0.50, 0.55, \ldots, 0.95\}$ in steps of $0.05$, and Recall1 is
reported at $\theta = 0.5$ and $\theta = 0.7$.
The exact matching and interpolation rules of $\mathrm{AP}_q(\theta)$ follow the
official evaluation script released with the baseline code.

Since this challenge focuses on how accurately the most confident retrieved
moment corresponds to a reference moment, the primary metric used for ranking is
\textbf{Recall1@0.7}, i.e., the top-ranked moment must attain $\IoU \geq 0.7$.
Note that participants are only required to predict a single moment, even when
multiple reference moments exist for a query, and submitting additional moments only affects mAP and does not change the primary ranking.

\section{Datasets}
\label{sec:dataset}
\begin{table}[t]
  \centering
  \caption{Overview of the datasets used in DCASE 2026 Challenge Task 6.
  \#Captions denotes the number of captions used as queries. Split
  assignments follow the redefined DCASE terminology described in
  \cref{ssec:devset}.}
  \label{tab:dataset_summary}
  \setlength{\tabcolsep}{4pt}
  \begin{tabular}{lcccc}
    \toprule
    Dataset & Annotation & \#Rec. & \#Captions & Length \\
    \midrule
    Clotho-Moment & synthetic   & 51{,}240 & 44{,}261 & 1\,min \\
    CASTELLA      & manual      & 1{,}862  & 3{,}881  & 1--5\,min \\
    Evaluation    & manual      & 100      & 177      & 1--5\,min \\
    \bottomrule
  \end{tabular}
  \vspace{-3mm}
\end{table}

For this task, we use two development datasets, Clotho-Moment~\cite{munakata2025amr} and
CASTELLA~\cite{castella}, and a separately collected evaluation dataset.
Each dataset consists of long audio recordings, free-form English captions, and the timestamps corresponding to each description.
To avoid issues related to downloading audio, the organizers also distribute pre-extracted MS-CLAP audio-text features of all three datasets, so that participants can train and evaluate systems without downloading the raw audio.
\cref{tab:dataset_summary} summarizes the datasets.

\subsection{Development dataset}
\label{ssec:devset}

\textbf{Clotho-Moment} is a large-scale synthetic dataset designed to boost the
training of AMR models~\cite{munakata2025amr}.
It is constructed by overlaying foreground audio clips from Clotho~\cite{clotho},
an audio-captioning dataset, onto long background recordings from the Walking
Tours videos~\cite{walkingtours} at random intervals drawn from an exponential
distribution with a mean of 30 seconds.
Silence at the onset and offset of each Clotho clip is removed before overlaying to clarify the start and end of the moments.
The foreground and background are mixed at randomly sampled relative levels.
Because the timestamps are obtained automatically from the overlay
procedure, no manual annotation is required, and the dataset is large, containing $51{,}240$ one-minute recordings with $44{,}261$ captions.

\textbf{CASTELLA} is a manually annotated dataset for training and
evaluation~\cite{castella}.
It contains $1{,}862$ recordings over $120$ hours in total.
This dataset is split into training, validation, and test splits, containing $2{,}182$, $352$, and $1{,}347$ annotated captions, respectively.
Recordings range from one to five minutes and is annotated with up to five moments.
The recordings were collected from the original long-form YouTube videos corresponding to a subset of AudioCaps.
The collected recordings shorter than one minute were excluded and longer ones were trimmed to five minutes.
The captions and one-second-resolution timestamps were collected via crowdsourcing. 
Annotators selected the salient moments of each recording and wrote their captions, and the annotations were further reviewed by listening to the audio only, so as to remove errors caused by visual bias.
On average, each recording contains $2.1$ moments, and each caption is $7.8$ words long.

CASTELLA and Clotho-Moment are originally divided into training, validation, and
test splits.
To ensure consistency, the split names are redefined using DCASE terminology:
\emph{development-training} comprises the training splits of both CASTELLA and Clotho-Moment;
\emph{development-validation} is the
CASTELLA validation split; and \emph{development-testing} is the CASTELLA test split.

\subsection{Evaluation dataset}
\label{ssec:evalset}

The evaluation dataset consists of $100$ real audio recordings collected from
YouTube, ranging from one to five minutes, with $177$ text queries in total.
The audio captions and timestamps were collected in the same manner as CASTELLA.
Each query may be associated with one or more ground-truth timestamp pairs, which were withheld during the challenge.

\section{Baseline system}
\label{sec:baseline}

The organizers provide a baseline system, AM-DETR~\cite{munakata2025amr}, that is simple but effective for this challenge.
It consists of two modules: an audio-text feature extractor and a
moment-detection network that adapts the video-moment-retrieval model
QD-DETR~\cite{qddetr} to the acoustic domain. QD-DETR builds on the Detection
Transformer (DETR)~\cite{detr} and Moment-DETR~\cite{momentdetr}.

\textbf{Feature extraction.}
The feature extractor is based on the pre-trained MS-CLAP 2023 model~\cite{msclap},
which maps audio and text into a shared cross-modal embedding space.
The text query is encoded by the text encoder into a sequence of 768-dimensional text embeddings. The audio recording is split using a sliding window with a 1-second window and a 1-second hop, and each frame is encoded by the audio encoder into a 768-dimensional audio embedding~\cite{munakata2025amr}.
Because both modalities are embedded in the same space, the resulting embeddings
already capture the alignment between acoustic content and language.
Operating on these pre-computed embeddings rather than on raw audio keeps the
detection network lightweight.

\textbf{Moment-detection network.}
Following QD-DETR~\cite{qddetr}, a transformer encoder contextualizes the sequence
of audio embeddings conditioned on the text-query embedding, and a transformer
decoder with $K$ learnable moment queries predicts a set of $K$ candidate moments parameterized by a center and width with a confidence score.
The default value of $K$ is 10 in the baseline setup.
At inference time, the candidate moments are ranked by their confidence scores,
and the most confident one is used for the primary Recall1@0.7 metric.

\textbf{Training and reproduction.}
The network is trained with bipartite matching between predicted and reference
moments. The training objective is a moment loss that combines an $L_1$
loss on the predicted center and width with a generalized IoU (gIoU)
loss~\cite{giou}, and a cross-entropy score loss on the
confidence, weighted by $\lambda_{L_1}$, $\lambda_{\mathrm{gIoU}}$, and
$\lambda_{\mathrm{score}}$, respectively. 
In addition, the baseline uses loss functions from QD-DETR~\cite{qddetr}, such as the highlight detection loss and the negative pair loss, enhancing the retrieval performance.
To make the setup reproducible, we state the protocol explicitly: models are
trained on the development-training split, selected on development-validation, and
evaluated on development-testing (\cref{ssec:devset}), and two configurations are
provided: one trained on CASTELLA only and one trained on CASTELLA together with
Clotho-Moment.
The baseline uses a two-layer transformer encoder and decoder with a hidden size of $256$, eight attention heads, and a feed-forward size of $1{,}024$, and is optimized with AdamW (learning rate $10^{-4}$, a batch size of $32$, for $200$ epochs).
The baseline is implemented based on the Lighthouse library~\cite{lighthouse}, and the full training and evaluation recipe has been released so that the reported numbers can be reproduced\footnote{The baseline code is available from~\url{https://github.com/awkrail/dcase2026_task6_baseline}}.

\section{Results and discussion}
\label{sec:experiments}

This section summarizes the challenge outcome and discusses the trends across the submitted systems.
\subsection{Results}
The challenge attracted 21 teams, which submitted 59 systems in total.
Each team was allowed up to four submissions, and the submissions were evaluated on the evaluation dataset of 100 real recordings and 177 queries (\cref{ssec:evalset}).
Table~\ref{tab:challenge_results} shows the performance and ranking of the best system for each team.
The top three systems tied on the primary metric, each achieving Recall1@0.7 of
$48.59\%$~\cite{Kibata_YCU_t6,Kim_CAU_t6,Sugawara_YCU_t6}, with the fourth-ranked
system achieving $46.89\%$~\cite{Ogawa_YCU_t6}.
These scores are roughly 3.5 times the baseline score.
The three tied systems are distinguished by the other metrics.
While \texttt{Kibata(YCU)} and \texttt{Kim(CAU)} also achieved strong results in Recall1@0.5 and average mAP, \texttt{Sugawara(YCU)} achieved a lower Recall1@0.5 score.
How these differences arise from the system designs is discussed in the following subsection.

\begin{table}[t]
  \centering
  \caption{Best system per team on the evaluation dataset, ranked by the primary
  metric Recall1@0.7 (\%). R1@$\theta$ denotes Recall1@$\theta$ and mAP is averaged
  over IoU $0.5\!:\!0.95$; higher is better. Tied systems are listed in alphabetical order.
  The values shown in gray in the mAP column are calculated using fewer than 10 predicted moments. * indicates a minor modification.}
  \label{tab:challenge_results}
  \scalebox{0.79}{
  \setlength{\tabcolsep}{4pt}
  \begin{tabular}{ll|cc|ccc}
    \toprule
    & & \multicolumn{2}{c|}{Architecture}\\
    \# & Team (institution) & Feature & Detection & R1@0.7 & R1@0.5 & mAP  \\
    & & extractor & network\\
    \midrule
    1  & Kibata (YCU)~\cite{Kibata_YCU_t6}            & M2D-CLAP & CG-DETR$^*$ & \textbf{48.59} & \textbf{69.49} & 40.67 \\
    1  & Kim (CAU)~\cite{Kim_CAU_t6}                  & Multiple & Original & \textbf{48.59} & 63.84 & \textbf{41.39} \\
    1  & Sugawara (YCU)~\cite{Sugawara_YCU_t6}        & Multiple & UVCOM & \textbf{48.59} & 59.89 & \textcolor{lightgray}{22.69} \\
    4  & Ogawa (YCU)~\cite{Ogawa_YCU_t6}              & M2D-CLAP & CG-DETR$^*$ & 46.89 & 60.45 & 34.59 \\
    5  & Calvet (AUDIAS)~\cite{Calvet_AUDIAS_t6}      & Multiple & UVCOM$^*$ & 43.50 & 64.41 & 33.15 \\
    6  & Choi (KAIST)~\cite{Choi_KAIST_t6}            & M2D-CLAP$^*$  & QD-DETR$*$ & 41.24 & 55.93 & 33.11 \\
    6  & Usui (YCU)~\cite{Usui_YCU_t6}                & M2D-CLAP$^*$ & UVCOM & 41.24 & 58.19 & 34.01 \\
    8  & Nakazawa (AM)~\cite{Nakazawa_AM_t6}          & Multiple & QD-DETR            & 36.16 & 49.72 & \textcolor{lightgray}{17.84} \\
    9  & Kang (ISCT)~\cite{Kang_ISCT_t6}              & FLAM & QD-DETR            & 35.59 & 50.28 & \textcolor{lightgray}{26.84} \\
    10 & Xiao (HEU)~\cite{Xiao_HEU_t6}                & MS-CLAP            & UVCOM$^*$ & 33.90 & 48.02 & 27.18 \\
    11 & Khan (WPI)~\cite{Khan_WPI_t6}                & Multiple            & 	UVCOM$^*$ & 32.20 & 48.02 & \textcolor{lightgray}{20.49} \\
    12 & Chunarkar (NTHU)~\cite{Chunarkar_NTHU_t6}    & Multiple & QD-DETR            & 29.38 & 52.54 & 22.73 \\
    13 & Huang (WHU)~\cite{Huang_WHU_t6}              & MS-CLAP            & UVCOM$^*$            & 27.68 & 48.02 & 23.58 \\
    14 & Nishijima (UTokyo)~\cite{Nishijima_UTokyo_t6}& Qwen2-Audio$^*$ & Qwen2-Audio & 22.60 & 32.77 & \textcolor{lightgray}{15.48} \\
    15 & Chen (CHT)~\cite{Chen_CHT_t6}                & MS-CLAP            & UVCOM & 22.03 & 37.29 & \textcolor{lightgray}{12.06} \\
    16 & Lu (YZU)~\cite{Lu_YZU_t6}                    & MS-CLAP            & QD-DETR            & 21.47 & 31.64 & 17.35 \\
    17 & Xu (GZHU)~\cite{Xu_GZHU_t6}                  & MS-CLAP            & QD-DETR$^*$            & 15.25 & 31.07 & 13.42 \\
    18 & Huck (NV)~\cite{Huck_NV_t6}                  & MS-CLAP$^*$ & - & 13.56 & 22.03 & 9.40 \\
    18 & Zhang (XJTLU)~\cite{Zhang_XJTLU_t6}          & MS-CLAP            & Original & 13.56 & 22.03 & 14.16 \\
    20 & Kret (Cooper Union)~\cite{Kret_CooperUnion_t6} & MS-CLAP          & - & 11.30 & 19.21 & \textcolor{lightgray}{6.57} \\
    21 & Minh (VGU)~\cite{Minh_VGU_t6}                & MS-CLAP            & Original & 5.65 & 12.99 & 6.89 \\
    \midrule
    -- & Baseline  & MS-CLAP & QD-DETR & 13.56 & 28.25 & 12.11 \\
    \bottomrule
  \end{tabular}
  }
\end{table}

\subsection{Discussion}
\label{ssec:discussion}

\textbf{Top-ranked systems.}
All three top-ranked teams~\cite{Kibata_YCU_t6, Sugawara_YCU_t6, Kim_CAU_t6} adopted approaches to refine the baseline.
While all three teams comprehensively improved the feature extractor, moment-detection network, loss functions, and post-processing techniques such as confidence score refinement, they focused on different components. \texttt{Kim(CAU)} and \texttt{Kibata(YCU)} focused on making various architectural improvements to enhance overall retrieval performance. 
\texttt{Sugawara(YCU)} focused on accurately estimating moment widths rather than moment centers through model ensembling.
This difference is reflected in the gap between Recall1@0.5 and Recall1@0.7,
suggesting that \texttt{Sugawara(YCU)} predicted more precise timestamps for
easier-to-detect moments, whereas \texttt{Kim(CAU)} and \texttt{Kibata(YCU)} were better at roughly detecting more difficult moments.

\textbf{Feature extraction.}
Table~\ref{tab:challenge_results} shows that the teams that replaced the feature extractor dominated the leaderboard, and their median Recall1@0.7 was about 41\%.
These teams adopted M2D-CLAP~\cite{m2dclap}, LAION-CLAP~\cite{laionclap}, FLAM~\cite{flam}, or self-supervised models like EAT and BEATs~\cite{eat,beats}.
\texttt{Kibata(YCU)} reported that simply replacing MS-CLAP features with M2D-CLAP features while keeping the architecture unchanged raised Recall1@0.7 from 26.80\% to 40.70\% on the development-testing dataset.
In contrast, systems that kept the provided MS-CLAP features achieved lower team-best scores, with a median Recall1@0.7  of 18\%.
However, a substantial gap remained between the top-ranked teams and teams such as \texttt{Nakazawa(AM)} that focused on exploring feature extractors~\cite{Nakazawa_AM_t6}, indicating that there is potential for further gains beyond just the feature extractor.

\textbf{Moment-detection network.}
Most systems used DETR-based moment-detection networks, often replacing QD-DETR with stronger variants.
Top-tier teams like~\cite{Kibata_YCU_t6,Sugawara_YCU_t6, Ogawa_YCU_t6, Calvet_AUDIAS_t6, Usui_YCU_t6} utilized existing powerful networks such as UVCOM~\cite{uvcom}, CG-DETR~\cite{cgdetr}, and BAM-DETR~\cite{bamdetr}.
\texttt{Kim(CAU)} and \texttt{Choi(KAIST)} independently replaced the transformer self-attention with a Mamba~\cite{mamba} temporal encoder and reported this replacement as one of their largest single-component gains~\cite{Kim_CAU_t6,Choi_KAIST_t6}.
In contrast, most systems based on a training-free approach did not outperform the baseline system, suggesting that a detection network trained on real data is essential for improving retrieval performance.

\begin{figure}[t]
    \centering
    \centering
    \begin{minipage}[t]{0.45\columnwidth}
        \centering
        \includegraphics[width=\columnwidth]{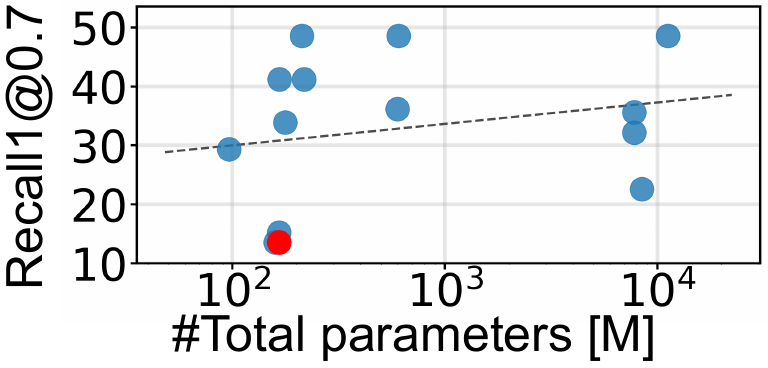}
    \end{minipage}
    \hfill
    \begin{minipage}[t]{0.45\columnwidth}
        \centering
        \includegraphics[width=\columnwidth]{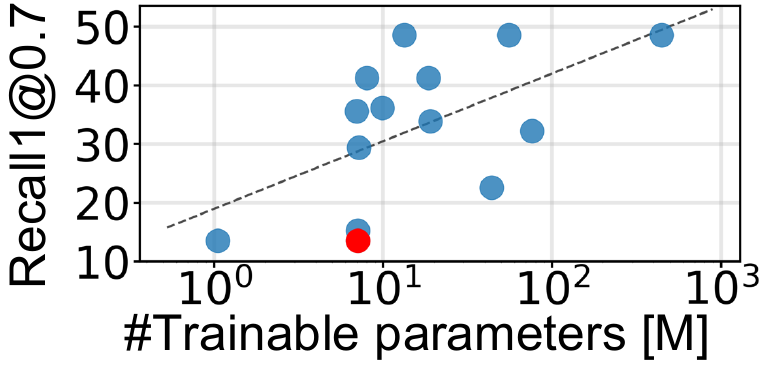}
    \end{minipage}
    \vspace{-2mm}
    \caption{Comparison of the retrieval performance on the evaluation dataset and the number of total or trainable parameters. In this figure, systems with invalid metadata were excluded.}
    \label{fig:model_params}
    \vspace{-2mm}
\end{figure}

\textbf{Model parameters.}
Fig.~\ref{fig:model_params} shows the relationship between the retrieval performance and the number of parameters.
The performance and the total number of parameters are only weakly correlated ($r=0.22$). Note that the systems with more than $10^4$ M parameters use audio LLMs.
The performance and the number of trainable parameters are more strongly correlated ($r=0.57$), whereas we can observe a system that achieves high performance without using more than ten times as many parameters as the baseline. 
These results imply that performance gains depend not only on scaling model parameters, but also on adding appropriate loss functions and incorporating post-processing methods such as confidence-score refinement and model ensembling.

\textbf{Confidence score refinement.}
Several teams addressed the misalignment between confidence scores and localization quality, where accurately localized moments could receive lower confidence scores than inaccurate ones.
Approaches to this problem can be broadly divided into two categories: adding a loss function and introducing a score calibrator.
For the loss function, IoU-aware approaches like varifocal loss~\cite{varifocal} were widely introduced.
\texttt{Usui(YCU)} reported that simply adding varifocal loss significantly improved Recall1@0.7 from 20.30\% to 27.62\%.
Score calibrators were implemented in various forms, such as adding a lightweight branch, constructing unique dedicated models, and leveraging Audio LLMs.
Two teams, including \texttt{Kim(CAU)}, added a branch for the score calibration to the network to directly predict IoU values during training, and the predicted IoU values were used for the calibration at inference time~\cite{Kim_CAU_t6, Huang_WHU_t6}.
Two other teams employed their own score calibrators~\cite{Calvet_AUDIAS_t6, Zhang_XJTLU_t6}. Notably, \texttt{Calvet(AUDIAS)} reported that an MLP-based calibrator improved Recall1@0.7 from 34.45\% to 40.01\%.
\texttt{Khan(WPI)} used an external model trained to be specialized in calibration~\cite{Khan_WPI_t6}, while others directly used models without any additional training~\cite{Kim_CAU_t6}.

\textbf{Timestamp refinement.}
There were also approaches that focused on refining timestamps rather than confidence scores.
\texttt{Sugawara(YCU)} reported that while detection networks can correctly predict the center of moments, they have a tendency to misestimate their widths~\cite{Sugawara_YCU_t6}.
To achieve accurate timestamp prediction, this team employed a model ensemble based on weighted box fusion~\cite{wbf}, specifically incorporating models with varied temporal resolutions of acoustic features.
This team tied for the highest Recall1@0.7 but had a lower Recall1@0.5 than the other two top teams, leaving a smaller gap between the two metrics.
This is likely because their approach improved the ability to estimate moment widths, but did not significantly improve the ability to predict moment centers.
Although \texttt{Kang(ISCT)} reported that a large-scale ensemble using models with the same architecture led to significant performance improvements~\cite{Kang_ISCT_t6}, the gap between their Recall1@0.5 and Recall1@0.7 scores does not show a clear trend compared to teams with similar rankings.
This result suggests that rather than relying on a simple ensemble, considering temporal resolution is crucial for improving the accuracy of moment-width estimation.

\textbf{Audio LLMs.}
While some teams exploited audio LLMs for the score refinement and data augmentation, \texttt{Nishijima(UTokyo)} leveraged Qwen2-Audio~\cite{qwen2audio} to directly predict the moments~\cite{Nishijima_UTokyo_t6}.
Since Qwen2-Audio cannot directly handle the long audio in AMR, this system compressed features along the sequence dimension and fine-tuned the model for AMR.
Although this system substantially outperformed the baseline in terms of Recall1@0.7, the gap with the top system remains significant.

\textbf{Synthetic-to-real gap.}
Several teams observed that systems tuned only on the synthetic development data often failed to transfer because of a mismatch between synthetic and real data. 
Indeed, when validation was performed on synthetic data, improvements in the validation score occasionally corresponded to worse performance on real data, so validating on the real
development-testing split proved essential.
To better exploit synthetic data, \texttt{Ogawa(YCU)} and \texttt{Kang(ISCT)} adopted data augmentation during pre-training.
These teams narrowed the gap with concatenation-based synthesis and mixing augmentations such as MomentMix~\cite{momentmix}, using external corpora~\cite{wavcaps}.

\begin{figure}[t]
    \centering
    \begin{minipage}[t]{0.32\columnwidth}
        \centering
        \includegraphics[width=\columnwidth]{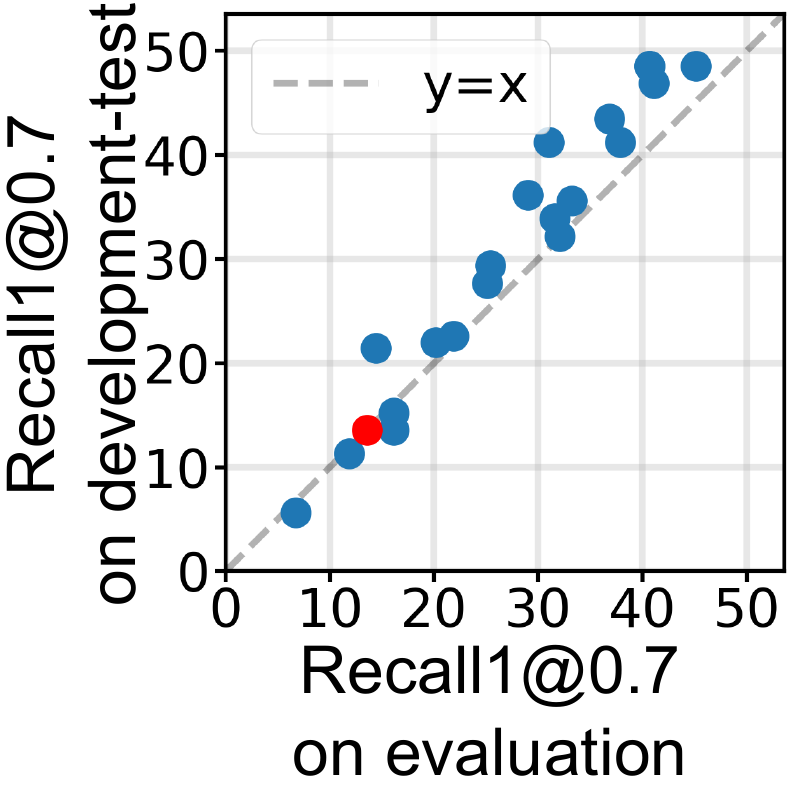}
    \end{minipage}
    \hfill
    \begin{minipage}[t]{0.32\columnwidth}
        \centering
        \includegraphics[width=\columnwidth]{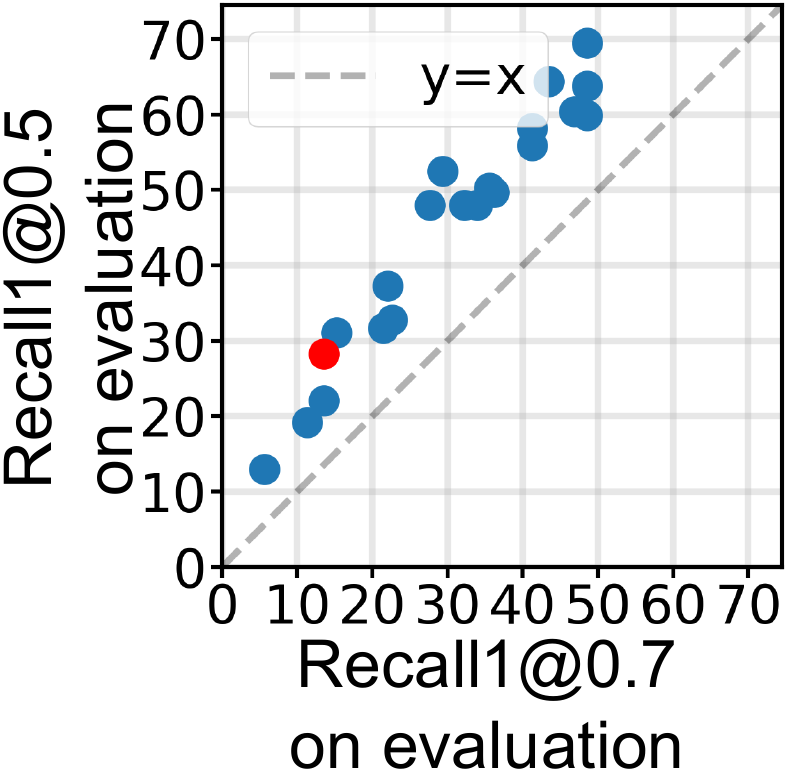}
    \end{minipage}
    \hfill
    \begin{minipage}[t]{0.32\columnwidth}
        \centering
        \includegraphics[width=\columnwidth]{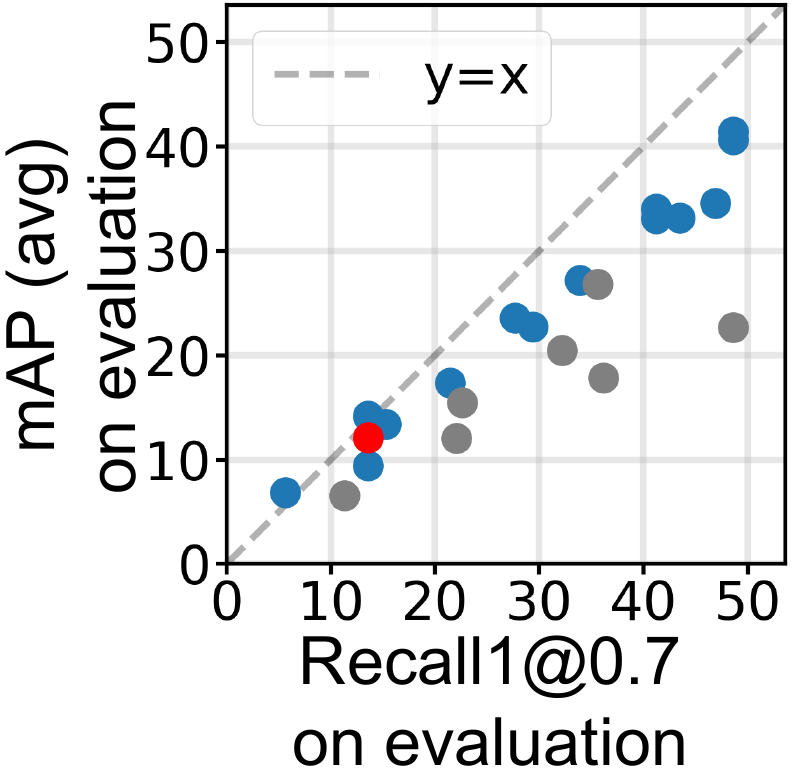}
    \end{minipage}
    \vspace{-2mm}
    \caption{Recall1@0.7 on the evaluation dataset and other metrics. The red dot in each graph represents the baseline system. The gray dots in the right graph represent the systems that submitted fewer than 10 predicted moments per query.}
    \label{fig:scatter}
    \vspace{-2mm}

\end{figure}

\textbf{Development-testing-to-evaluation gap.}
The left panel of Fig.~\ref{fig:scatter} demonstrates the correlation between Recall1@0.7 on the development-testing dataset and that on the evaluation dataset. While both datasets were collected from YouTube, the development-testing dataset was obtained from a subset of AudioCaps, and the distribution of moments reflects the bias caused by AudioCaps' filtering process.
However, the very strong correlation between these scores suggests that the domain mismatch between the development-testing and evaluation datasets is limited.

\textbf{Comparison between metrics.}
The center and right graphs of Fig.~\ref{fig:scatter} show relationships between metrics on the evaluation dataset.
Comparing Recall1@0.7 and Recall1@0.5, we can find that the gap between these metrics widens as the score increases, moving further away from the $y=x$ line.
In light of the discussion on timestamp refinement, systems with a large gap between the two metrics are likely to achieve further gains in Recall1@0.7 if they improve their performance in predicting moment widths.
Comparing Recall1@0.7 and average mAP, the gap between these metrics also widens as the scores increase. This indicates that predicting multiple intervals remains challenging.

\section{Conclusion}
\label{sec:conclusion}

This paper describes DCASE 2026 Challenge Task 6, Audio Moment Retrieval from Long Audio, the first DCASE task to address the retrieval of text-queried temporal moments from minutes-long recordings.
We provide the datasets, the baseline model and the evaluation protocol.
For the evaluation, we constructed a manually annotated evaluation dataset dedicated to this challenge.
The baseline system combines an MS-CLAP feature extractor with a DETR-based moment-detection network.
The systems are evaluated using the IoU-based ranking metrics with Recall1@0.7 as the primary metric.
The challenge drew 21 teams and 59 systems, and the three best-performing systems achieved a Recall1@0.7 of $48.59\%$, roughly 3.5 times the baseline score.
All the top-performing systems enhanced their feature extractors and moment-detection networks.
In particular, the three top-ranked systems incorporated further ideas, such as introducing a score calibrator and ensembling models with varying temporal resolutions of features. The gap between Recall1@0.5 and Recall1@0.7 suggests that there is still room for further performance improvement. Future work will focus on constructing a simple yet higher-performing model by integrating and refining these approaches.

\section{Acknowledgments}
We thank Paul Primus from Johannes Kepler University Linz for his invaluable support in preparing the challenge.

\clearpage
\bibliographystyle{IEEEtran}
\bibliography{refs}

\end{document}